\documentclass[prb,twocolumn,showpacs,preprintnumbers,amsmath,amssymb,superscriptaddress,nofootinbib]{revtex4}

\usepackage{bbm,bm,slashed}
\usepackage{graphicx,graphics,psfrag,epsfig}
\usepackage{color,array,tabularx}
\usepackage{hyperref}

\begin{document}

\title{Interacting electrons on trilayer honeycomb lattices} 

\date{\today} 

\author{Michael M. Scherer}
\email{scherer@physik.rwth-aachen.de}
\affiliation{Institute for Theoretical Solid State Physics,  RWTH Aachen University, Germany}
\affiliation{JARA-FIT Fundamentals of Future Information Technology}
\author{Stefan Uebelacker}
\affiliation{Institute for Theoretical Solid State Physics, RWTH Aachen University, Germany} 
\affiliation{JARA-FIT Fundamentals of Future Information Technology}
\author{Daniel D. Scherer}
\affiliation{Theoretisch-Physikalisches Institut, Friedrich-Schiller-Universit\"at Jena, Germany} 
\author{Carsten Honerkamp}
\affiliation{Institute for Theoretical Solid State Physics, RWTH Aachen University, Germany} 
\affiliation{JARA-FIT Fundamentals of Future Information Technology}


\begin{abstract} 
Few-layer graphene systems come in various stacking orders. Considering tight-binding models for electrons on stacked honeycomb layers, this gives rise to a variety of low-energy band structures near the charge neutrality point. Depending on the stacking order these band structures enhance or reduce the role of electron-electron interactions.  Here, we investigate the instabilities of interacting electrons on honeycomb multilayers with a focus on trilayers with ABA and ABC stackings theoretically by means of the functional renormalization group. We find different types of competing instabilities and identify the leading ordering tendencies in the different regions of the phase diagram for a range of local and non-local short-ranged interactions. The dominant instabilities turn out to be toward an antiferromagnetic spin-density wave (SDW), a charge density wave and toward quantum spin Hall (QSH) order. Ab-initio values for the interaction parameters put the systems at the border between SDW and QSH regimes.
Furthermore, we  discuss the energy scales for the interaction-induced gaps of this model study and put them into context with the scales for single-layer and Bernal-stacked bilayer honeycomb lattices. This yields a comprehensive picture of the possible interaction-induced ground states of few-layer graphene.
\end{abstract}

\maketitle



\section{Introduction}

In the last years, the electronic properties of few-layer graphene systems have been studied extensively \cite{castroneto2009}. Experiments on graphene systems with two\cite{exp01,exp02,exp03,exp04,exp05,exp06,exp07,exp08} and three layers\cite{bao2010,taychatanapat2011,lyzhang2011,kumar2011,jhang2011,lui2011,eisenstein2011,russo2011} aimed to clarify the role of many-body interactions\cite{bao2011,norismatsu}. These are expected to be more effective due to the enhanced low-energy density of states and thus correlated electronic ground states may be realized. For bilayer graphene the experimental studies agree on the correlated nature of the ground state, but disagree on the symmetries of the underlying order, a topic that is matter of current debate. At temperatures below 5K, transport experiments found gap openings of about 2-3~meV \cite{exp06,exp07,exp08}. For trilayer graphene the stacking order is crucial for the electronic properties\cite{mccann2009,fzhang2010,avetisyan2010,barlas2011,gelderen2011,katsnelson2011,mccann2011,fzhang2012,jung2012,xu2012}. In Ref.~\onlinecite{bao2011}, it was found that neutral trilayer graphene with ABC stacking shows many-body correlations with a pronounced gap of $\sim 6\,$meV while graphene trilayers with ABA stacking do not show a gap. In view of this experimental situation a better theoretical understanding of the many-body instabilities and the nature of the correlated phases which are candidates for possible trilayer graphene ground states is required.

Here, we employ a functional renormalization group (fRG) approach (for a recent review, see Ref.~\onlinecite{metzner2011}) to address the problem of competing instabilities on honeycomb trilayers in an unbiased way. We explore a region of the phase diagram with a range of interaction parameters with density-density repulsions up to the second nearest neighbor that are motivated by the ab-initio values as proposed in Ref.~\onlinecite{wehling2011}. In a previous study, our RG approach was applied to the AB stacked honeycomb bilayer\cite{scherer2012} to investigate the phase diagram as a function of local and non-local interaction parameters and complemented by a more quantitative study for pure onsite interactions combining mean-field theory, functional RG and quantum Monte Carlo (QMC) techniques\cite{lang2012}. These studies clarified the relevance of various phases on the bilayer that have been subject to previous theoretical studies by different authors\cite{nilsson2006,mccann2007,hongki2008,lemonik2010,vafek2010a,nandkishore2010,zhang2010,jung2011,vafek2010b,
vafek2011,khari2011,kotov2010}, e.g. the antiferromagnetic spin density wave (AF-SDW), two kinds of charge density waves (CDW, CDW$_3$) and a quantum spin Hall state (QSH), depending on the model parameters when the full four-band model with the available ab-initio estimates for the model parameters is used. 
 The combination with QMC simulations\cite{lang2012} showed the compatibility of fRG and QMC methods for the investigated range of parameters, and allowed to gauge the quantitative precision of the fRG data at least partially.

The instability toward an interaction-induced quantum spin Hall (QSH) state is of particular interest in connection with realizing topological insulators in graphene. While the original proposal by Kane and Mele\cite{kanemele2005}, extending earlier work by Haldane\cite{haldane1988} with a mass term due to spin-orbit interactions turned out to too small gaps\cite{}, it was later argued\cite{HonerkampRaghu2008} that second-nearest neighbor repulsions could lead to a interaction-induced mean-field having the same effect as the Kane-Mele mass term. In the single graphene layer, however, a nonzero interaction strength is needed to open any kind of gap. So far there is no experimental evidence for this. In the bilayer system, the same instability occurs for arbitrarily small interactions of appropriate distance dependence (see e.g. Ref.~\onlinecite{scherer2012}), and one basically gets two copies of the Kane-Mele QSH state coupled by the interlayer terms. The sign of the order parameter gives rise to distinct choices with interestingly different properties\cite{zhu2012}. While this is an interesting many-body state, the corresponding spin-polarized edge modes would not be topologically protected in a strict sense as the bilayer-doubling now permits time-reversal invariant single-particle terms that would localize the edge states\cite{qizhang}. However, if one finds the same state in the trilayer system, one would again have an odd number of Kramers pairs at the edges or edge states per spin, and the topological protection would keep alive at least one edge mode. In this work we show that the fRG suggests that the trilayer QSH state is not unlikely for realistic parameters, and that it can even occur at sizable energy scales $\sim 10\,$meV.

This paper is organized as follows. In Sec.~\ref{modelhamiltonians}, we introduce simple tight-binding Hamiltonians for ABC and ABA trilayer honeycomb lattices and discuss the resulting band structures depending on the stacking order. Further, we introduce the interacting part of the Hamiltonian with onsite, nearest and next-to-nearest neighbor interaction terms. In Sec.~\ref{frgmethod} we describe the fRG method and discuss the patching scheme as well as the employed approximations, together with the relations to other methods. In Sec.~\ref{frgresults} we analyze the emergent effective interactions from the fRG flow. We classify the instabilities and ordering tendencies in ABA and ABC trilayer honeycomb lattices for interacting electrons. With this information, we can construct the tentative phase diagram from fRG in the space of interaction parameters. We finish with a discussion of the energy scales for the gaps in the electronic spectrum due to the ordering phenomena in Sec.~\ref{energyscales} and some conclusions in Sec.~\ref{discussion}.


\section{Model Hamiltonians}
\label{modelhamiltonians}


\subsection{Basic Hamiltonians}

\begin{figure}[t!]
\centering
 \includegraphics[width=1.0\columnwidth]{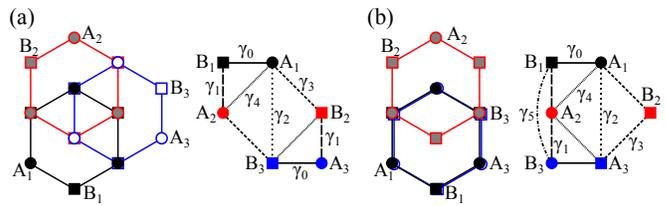}
 \caption{(a) Sketch of ABC stacked trilayer graphene with trigonal warping terms. (b)  Sketch of ABA stacked trilayer graphene with trigonal warping terms. The hoppings are shown in side-view.}
\label{fig:trilayer}
\end{figure}

\begin{figure}[t!]
\centering
 \includegraphics[width=0.49\columnwidth]{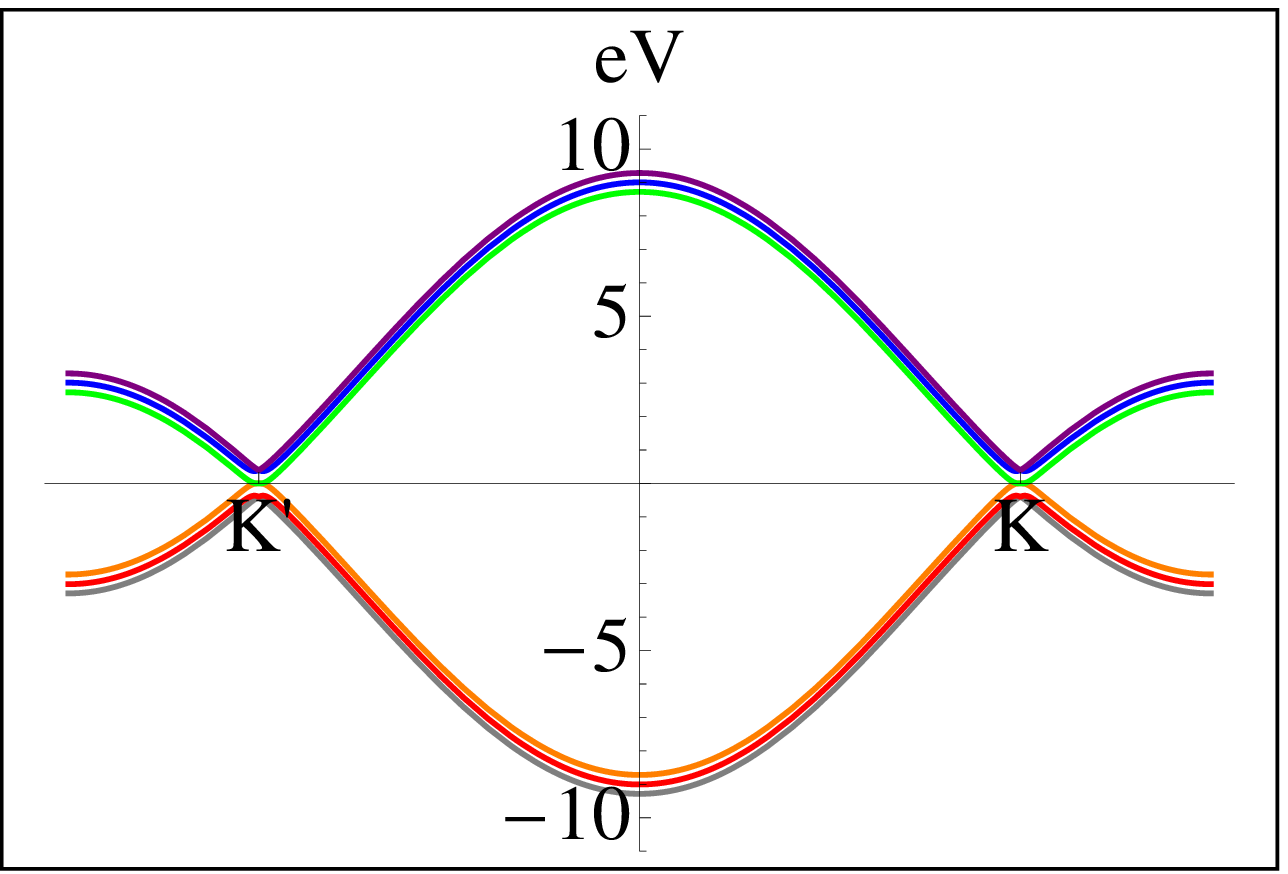}
 \includegraphics[width=0.49\columnwidth]{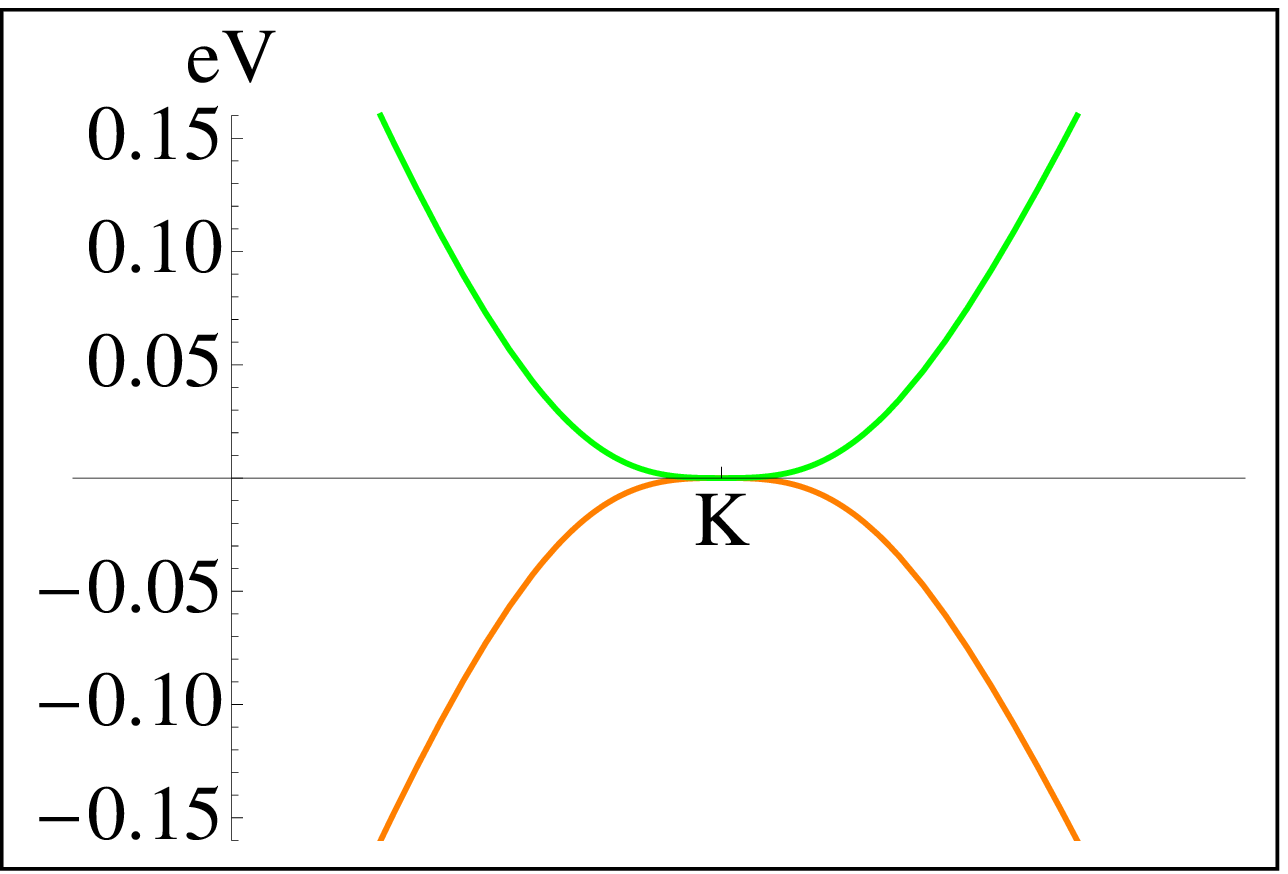}
\includegraphics[width=0.49\columnwidth]{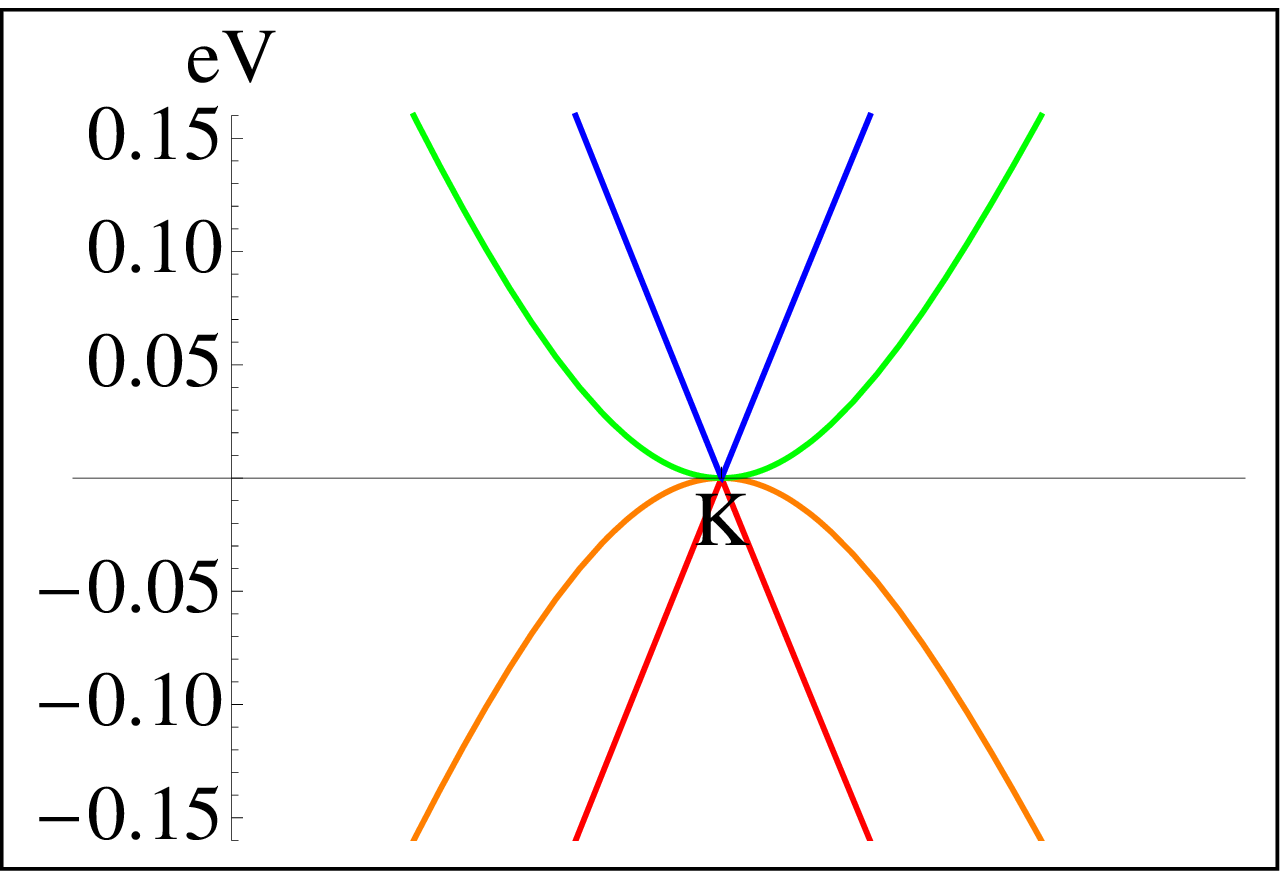}
\includegraphics[width=0.49\columnwidth]{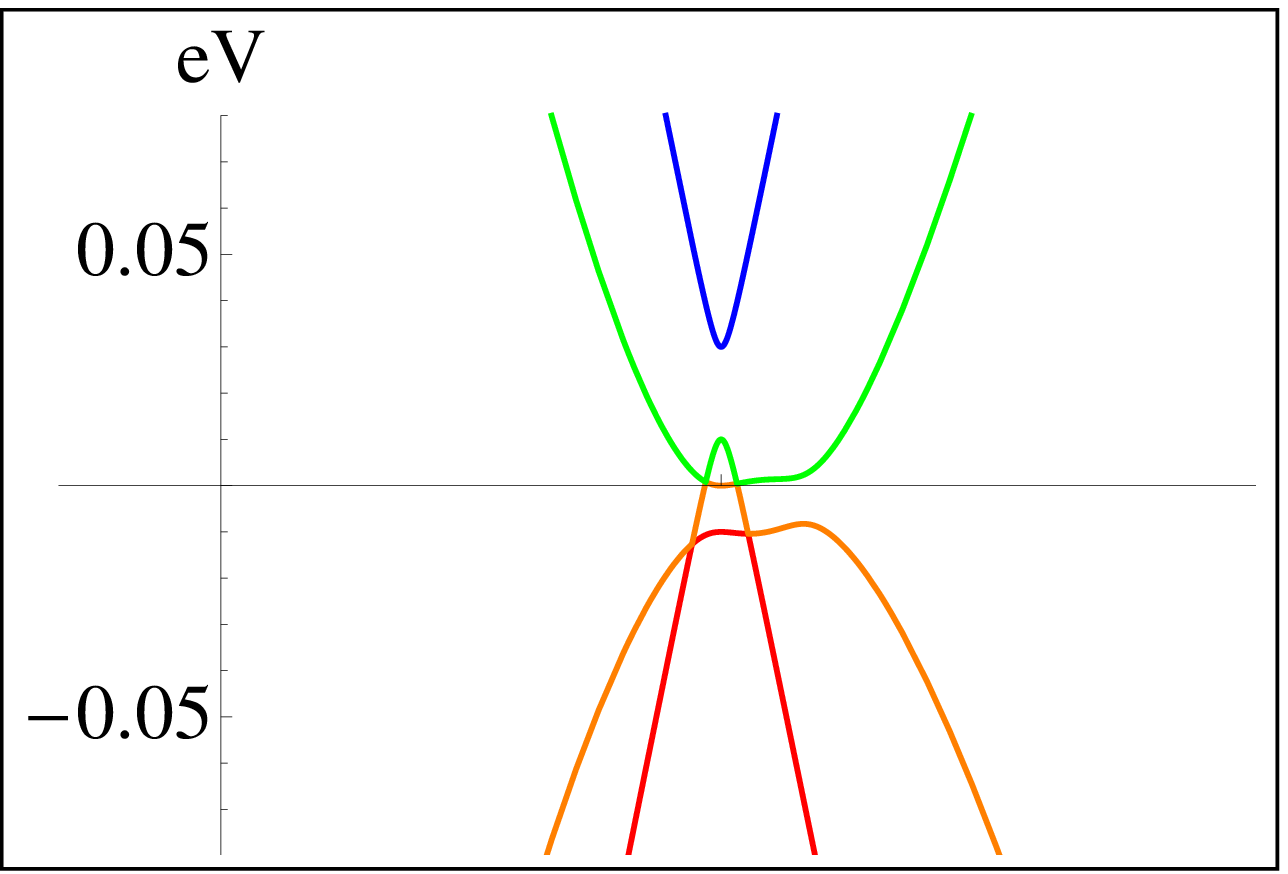}
\caption{Sketch of ABC and ABA stacked trilayer graphene band structure with $t=3\,$eV, $t_\perp=0.4\,$eV. Top left panel:  ABC trilayer dispersion with $k_y=0$. Top right panel: ABC trilayer dispersion close to $K, K'$ with $k_y=0$. Bottom left panel:  ABA bands close to $K, K'$ with $k_y=0$. Bottom right panel:  ABA bands close to $K, K'$ with $k_y=0$ with remote hoppings $\gamma_2, \gamma_3,\gamma_4, \gamma_5$ and onsite energy $\delta$ and the numerical values as given in the main text.}
\label{fig:trilayerdisp}
\end{figure}

Here, we construct the non-interacting part of the tight-binding Hamiltonians for ABC and ABA stacked trilayer honeycomb lattices. The position vectors of the two-dimensional bipartite lattice structure are called $\vec{R}$, with each $\vec{R}$ having three nearest neighbors at positions that can be characterized by the three vectors $\vec{\delta}_n$ with $n \in \{1,2,3\}$. 
Explicitly, the $\vec{\delta}_n$ are given by 
$\vec{\delta}_1= \frac{\sqrt{3}a}{2} \vec{e}_x+\frac{a}{2}\vec{e}_y$, $\vec{\delta}_2= -\frac{\sqrt{3}a}{2} \vec{e}_x+\frac{a}{2}\vec{e}_y$ and $\vec{\delta}_3= -a\vec{e}_y$.
Here, $a$ is the distance between two neighboring lattice sites and the vectors point from the B-sublattice to the A-sublattice. The tight-binding Hamiltonian for ABC/ABA stacked trilayers  is composed out of single layer Hamiltonians for the in-plane hoppings, perpendicular hoppings between different layers and remote hoppings between the layers including a certain planar distance. We denote the different lattice sites by $A_1, B_1,A_2, B_2,A_3, B_3$, where $A$ and $B$ specify the sublattice and the index numbers the layer as shown in Fig.~\ref{fig:trilayer}. 
Accordingly, we define the annihilation operators $c_{o,s,\vec{r}}$ and the creation operators $c_{o,s,\vec{r}}^\dagger$ of an electron at position $\vec{r}$ and the layer and sublattice are encoded in the subscript $o \in \{a_1,b_1,a_2,b_2,a_3,b_3\}$. The index $s= \uparrow,\downarrow$ specifies the electron spin. With these preliminaries we can write down the single-layer tight-binding Hamiltonians in the first two layers,
\begin{eqnarray}
H_{1}^{\parallel}&=&-\gamma_0\sum_{s,\vec{R}, \vec{\delta}_i}\Big(c_{b_1,s,\vec{R}}^\dagger c_{a_1,s,\vec{R}+\vec{\delta}_i}+\mathrm{h.c.}\Big),\\
H_{2}^{\parallel}&=&-\gamma_0\sum_{s,\vec{R},\vec{\delta}_i}\Big(c_{b_2,s,\vec{R}-\vec{\delta}_i}^\dagger c_{a_2,s,\vec{R}}+\mathrm{h.c.}\Big),
\end{eqnarray}
and a stacking dependent Hamiltonian for the third layer,
\begin{equation}
H_{3,c}^{\parallel}=-\gamma_0\sum_{s,\vec{R},\vec{\delta}_i}\Big(c_{b_3,s,\vec{R}+c\vec{\delta}_1}^\dagger c_{a_3,s,\vec{R}+c\vec{\delta}_1+\vec{\delta}_i}+\mathrm{h.c.}\Big)\,,
\end{equation}
where we have attributed an additional index $c$ to $H_{3,c}$, with $c=0$ for the ABA stacking and $c=1$ for the ABC or chiral stacking. In graphene and few-layer graphene systems the hopping $\gamma_0=t\approx 3\,$eV, see e.g. Refs.~\onlinecite{Zhang2008,zhang2010}.
Next we introduce the interlayer hoppings, $\gamma_1$, between sites that lie on top of each other and connect adjacent layers,
\begin{eqnarray}
H_{12}^{\perp}&=&\gamma_1\sum_{s,\vec{R}}\Big(c_{b_1,s,\vec{R}}^\dagger c_{a_2,s,\vec{R}}+\mathrm{h.c.}\Big)\,\\
H_{23,0}^{\perp}&=&\gamma_1\sum_{s,\vec{R}}\Big(c_{b_3,s,\vec{R}}^\dagger c_{a_2,s,\vec{R}}+\mathrm{h.c.}\Big)\,\\
H_{23,1}^{\perp}&=&\gamma_1\sum_{s,\vec{R}}\Big(c_{b_2,s,\vec{R}-\vec{\delta}_1}^\dagger c_{a_3,s,\vec{R}-\vec{\delta}_1}+\mathrm{h.c.}\Big)\,,
\end{eqnarray}
where again we accounted for the different stackings by introducing $H_{23,c}^\perp$. \emph{Ab initio} values for $\gamma_1=t_\perp$ are available for graphite\cite{dresselhaus2002,mccann2011}, $t_\perp\approx 0.4\,$eV , and ABC trilayer graphene\cite{zhang2010}, $t_\perp\approx 0.5\,$eV. 
In Fig.~\ref{fig:trilayer}, we also show the more remote hoppings $\gamma_2,\gamma_3,\gamma_4,\gamma_5$ whose effect will be discussed in the following section. In the present model study we will ignore these terms for most explicit calculations. For the discussion of the energy bands, 
we introduce the Fourier transform 
\begin{equation}
c_{o,s,\vec{k}}=\frac{1}{\sqrt{\mathcal{N}}}\sum_{\vec{r}}e^{i\vec{k}\cdot\vec{r}}c_{o,s,\vec{r}}\,,
\end{equation} 
where $\mathcal{N}$ is the number of unit cells and $\vec{k}$ is an element of the first Brillouin zone (BZ). We can write the tight-binding Hamiltonians in this approximation in Fourier space as
\begin{eqnarray}
H^{\mathrm{TL}}_c&=&H_{1}^{\parallel}+H_{2}^{\parallel}+H_{3,c}^{\parallel}+H_{12}^{\perp}+H_{23,c}^{\perp}\nonumber\\
&=&\sum_{s,\vec{k}}\varphi^\dagger_{s,\vec{k}}
\hat H^{\mathrm{TL}}_c
\varphi_{s,\vec{k}}\,,
\end{eqnarray}
with $\varphi^\dagger_{s, \vec{k}}:=(c^\dagger_{a_1,s,\vec{k}},c^\dagger_{b_1,s,\vec{k}},c^\dagger_{a_2,s,\vec{k}},c^\dagger_{b_2,s,\vec{k}},c^\dagger_{a_3,s,\vec{k}},c^\dagger_{b_3,s,\vec{k}})$ and
\begin{equation}
\hat H^{\mathrm{TL}}_c=\gamma_0
\begin{pmatrix}
0 &  -d_k & 0 & 0 & 0 & 0\\
 -d_k^\ast & 0 & \frac{\gamma_1}{\gamma_0} & 0 & 0 & 0\\
0 & \frac{\gamma_1}{\gamma_0} & 0 &  -d_k & 0 & \frac{(1-c)\gamma_1}{\gamma_0}\\
0 & 0 & -d_k^\ast & 0 & \frac{c\gamma_1}{\gamma_0} & 0\\
0 & 0 & 0 & \frac{c\gamma_1}{\gamma_0} & 0 & -d_k\\
0 & 0 &\frac{(1-c) \gamma_1}{\gamma_0} & 0 & -d_k^\ast & 0
\end{pmatrix}\,,
\end{equation}
where $d_k=\sum_n \exp(i \vec{k}\cdot \vec{\delta}_n)$. The resulting energy bands are depicted in Fig.~\ref{fig:trilayerdisp}. The dispersion for ABC trilayers is very flat near the Fermi level with cubic wavevector dependence close to the $K$ and $K^\prime$ points. Hence one should expect an enhanced role of interactions as compared to ABA trilayers and AB bilayers where the wavevector dependence is quadratic.


\subsection{Remote hoppings in trilayers}

In this section, we want to discuss the effect of remote hopping terms on the band dispersion and possible implications for instabilities. For the ABC trilayer system the density of states close to the Fermi energy has a van Hove singularity $\sim \epsilon^{-1/3}$ due to the cubic band crossing point. This enhances the role of interaction effects, leading to high critical scales for ordering tendencies. In our model study, we therefore choose to ignore remote hoppings that change the topology of the bands below an energy of order $10\,$meV \cite{zhang2010}, which is of the order of the measured gap in the experiment. Of course, it is an interesting question whether these terms affect the nature of the instability in trilayer graphene. However, we find in our computations that instabilities occur already on an energy scale that is higher than or at least comparable to $0.004t\sim 10\,$meV. This serves as an \textit{a posteriori} justification for dropping the remote hopping terms and proves the consistency of our approach.

In ABA trilayer the situation is different. Here, the band structure close to the $K$-, $K^\prime$-points is separated into a $J=1$ (linear) and a $J=2$ (quadratic) subband. The remote hoppings induce separate gaps for these subbands, however, with an individual energy shift of the subbands. This destroys the particle-hole nesting and the associated instabilities are suppressed (at least for small interaction terms). Explicitly, the tight-binding Hamiltonian in presence of the important remote hoppings\cite{dresselhaus2002,mccann2011,fzhang2012}, $\gamma_2= -0.02\ \mathrm{eV},\gamma_3= 0.3\ \mathrm{eV}, \gamma_4 = 0.04\ \mathrm{eV},\gamma_5= 0.04\ \mathrm{eV},\delta= 0.05\ \mathrm{eV}$, can be written as
\begin{equation}
\hat H^{\ast}
= 
\begin{pmatrix}
0 & -\gamma_0 d_k & \gamma_4 d_k & -\gamma_3 d_k^\ast & \frac{\gamma_2}{2} & 0\\
-\gamma_0 d_k^\ast & \delta & \gamma_1 & \gamma_4 d_k & 0 & \frac{\gamma_5}{2}\\
\gamma_4 d_k^\ast & \gamma_1 & \delta & -\gamma_0 d_k & \gamma_4 d_k^\ast & \gamma_1\\
-\gamma_3 d_k & \gamma_4 d_k^\ast & -\gamma_0 d_k^\ast & 0 & -\gamma_3 d_k & \gamma_4 d_k^\ast\\
\frac{\gamma_2}{2}  & 0 & \gamma_4 d_k & -\gamma_3 d_k^\ast & 0 & -\gamma_0 d_k\\
0 & \frac{\gamma_5}{2}  & \gamma_1 & \gamma_4 d_k & -\gamma_0 d_k^\ast & \delta \end{pmatrix}.
\end{equation}
Within the fRG approach, we will show that also for this system flows toward many-body instabilities occur, however, only beyond a critical interaction strength which, similarly to the single-layer case, most probably is larger than the interaction strength in the real material.


\subsection{Interaction terms} 

In order to investigate the instabilities that are possible on the trilayer honeycomb lattice for interacting electrons we will take into account a number of different interaction terms, most importantly an onsite or Hubbard interaction $U$, a nearest neighbor intralayer interaction $V_1$ and a next-to-nearest neighbor intralayer interaction $V_2$. For these interaction parameters ab initio parameters from constrained random phase approximation (cRPA) computations are available \cite{wehling2011} and we take those values as a motivation for the investigated range in the phase diagram. The interaction Hamiltonian reads
\begin{eqnarray}\label{eq:longH}
 H_\mathrm{I}&=&U\sum_i n_{i,\uparrow}n_{i,\downarrow}+V_1\sum
 _{\langle i,j \rangle, s, s^\prime} 
 n_{i,s}n_{j,s^\prime}\\
 &&+V_2\sum
 _{\langle \langle i,j \rangle \rangle, s, s^\prime} 
 n_{i,s}n_{j,s^\prime}\,,\nonumber
\end{eqnarray}
where $n_{i,s}=c^\dagger_{o,s,i}c_{o,s,i}$ and $i,j$ run over the lattice sites, but pairs are included only once. The unitary transformation from the orbital to the band degrees of freedom diagonalizing $H_0^{\mathrm{TL}}$ or  $H_1^{\mathrm{TL}}$ is performed numerically and has to be carried out on $H_\mathrm{I}$ as well. This adds `orbital makeup' to the interaction terms in band representation, leading to an additional angular dependence of the interactions near the $K$, $K'$-points. In order to resolve this dependence we use an angular patching of the interaction terms as explained in Sec.~\ref{frgmethod}.


\section{fRG method}
\label{frgmethod}

We employ a functional renormalization group (fRG) approach for the one-particle-irreducible vertices with a momentum cutoff (for a recent review on the fRG method, see \cite{metzner2011}). In this scheme, an infrared regulator with energy  scale $\Lambda$  is introduced into the bare propagator function,
\begin{equation}
G_0(\omega,\vec{k},b)\rightarrow G_0^\Lambda(\omega,\vec{k},b)=\frac{C^\Lambda [\epsilon(\vec{k},b)] }{i \omega -\epsilon(\vec{k},b) }.
\end{equation}
Here, $\omega$ is the Matsubara frequency, $\vec{k}$ the wavevector, $b$~the band index and $\epsilon(\vec{k},b)$ is the single-particle dispersion. As the spin is conserved the free propagator is spin independent. The cutoff function is chosen to enforce an energy cutoff, which regularizes the free Green's function by suppressing the modes with band energy below the scale $\Lambda$,
\begin{equation}
C^\Lambda [\epsilon(\vec{k},b)]  \approx \Theta \left( |\epsilon(\vec{k},b)|  - \Lambda \right)\,.
\end{equation}
For better numerical feasibility the step function is slightly softened in the actual implementation. With this modified scale-dependent propagator, we can define the scale-dependent effective action $\Gamma_\Lambda$ as the Legendre transform of the generating functional $\mathcal G_\Lambda$ for correlation functions, cf. Ref.~\onlinecite{negorl}.
The RG flow of $\Gamma_\Lambda$ is generated upon variation of $\Lambda$. By integrating the flow down from an initial scale $\Lambda_0$, which is in our case chosen as the maximum energy of all bands, to the infrared $\Lambda \rightarrow 0$, one smoothly interpolates between the bare action of the system and the effective action at low energy. In order to limit the numerical effort we employ the following approximations. First, the hierarchy of flowing vertex functions is truncated after the four-point (two-particle interaction) vertex. Secondly we neglect the frequency dependence, by setting all external frequencies to zero, as we are interested in ground-state properties. The general coupling function, which depends on three momentum and Matsubara frequency indices (the fourth index is fixed by conservation) is thus replaced by the coupling function $V^\Lambda(k_1,k_2,k_3,b_4)$ in band representation with $k_i=(\vec{k}_i,b_i)$ or in orbital representation by $V^\Lambda(\tilde k_1,\tilde k_2,\tilde k_3,o_4)$ with $\tilde k_i=(\vec{k}_i,o_i)$. Third, we neglect self-energy corrections. This approximate fRG scheme then amounts to an infinite-order summation of one-loop particle-particle and particle-hole terms of second order in the effective interactions. It allows for an unbiased investigation of the competition between various correlations, by analyzing the components of $V^\Lambda(\tilde k_1,\tilde k_2,\tilde k_3,o_4)$ that create instabilities by growing large at a critical scale $\Lambda_c$\cite{metzner2011}. With the approximations mentioned above, this procedure is well-controlled for small interactions. At intermediate interaction strengths we still expect to obtain reasonable results. In any case, the fRG takes into account effects beyond mean-field and random phase approximations.

The fRG calculation is performed in the band basis in which the free part of the Hamiltonian is diagonal. The free Hamiltonian can be diagonalized by a unitary transformation of the form
\begin{eqnarray}
c_{b,s,\vec{k}}&=&\sum_o u_{bo,\vec{k}} c_{o,s,\vec{k}},\\
c^{\dagger}_{b,s,\vec{k}}&=&\sum_o u^{*}_{bo,\vec{k}} c^{\dagger}_{o,s,\vec{k}}\,,
\end{eqnarray}
where the index $o$ includes all six sublattices in the three layers and the index $b$ denotes the corresponding bands. 
In the interaction part of the Hamiltonian the coupling function $V^\Lambda (\tilde k_1,\tilde k_2,\tilde k_3,o_4) $ in orbital space is chosen so that it reproduces the interaction Hamiltonian of Eq.~ \eqref{eq:longH}. 
For application of the fRG in the band basis we have to transform the interaction via
\begin{eqnarray}\label{eq-om}
\label{omakeup}
H_{\mathrm I} &=& \frac{1}{2 \, \mathcal{N}} \sum_{\vec{k}_1, \vec{k}_2, \vec{k}_3, s, s' \atop o_1, o_2, o_3, o_4} 
V (\tilde k_1,\tilde k_2,\tilde k_3,o_4) \nonumber \\[1mm] && \times \, \sum_{b_1,b_2,b_3,b_4}
u_{o_1b_1,\vec{k}_1} u_{o_2b_2,\vec{k}_2}u^*_{o_3b_3,\vec{k}_3}u^*_{o_4b_4,\vec{k}_4}
\nonumber \\[1mm] && \times \,
c^{\dagger}_{b_3,s,\vec{k}_3} c^{\dagger}_{b_4,s',\vec{k}_4} c_{b_2,s',\vec{k}_2} c_{b_1,s,\vec{k}_1} \nonumber \\
&=& \frac{1}{2 \, \mathcal{N}} \sum_{\vec{k}_1, \vec{k}_2, \vec{k}_3, s, s' \atop b_1, b_2,b_3, b_4} V(k_1,k_2,k_3,b_4)\nonumber \\ && \times \, 
c^{\dagger}_{b_3,s,\vec{k}_3} c^{\dagger}_{b_4,s',\vec{k}_4} c_{b_2,s',\vec{k}_2} c_{b_1,s,\vec{k}_1} 
\, . 
\end{eqnarray}
Thus, in the band basis the interaction vertex $V(k_1,k_2,k_3,b_4)$ acquires a pronounced momentum dependence due to the extra prefactor $u_{o_1b_1,\vec{k}_1} u_{o_2b_2,\vec{k}_2}u^*_{o_3b_3,\vec{k}_3}u^*_{o_4b_4,\vec{k}_4}$, which is sometimes also referred to as `orbital makeup'. Previous studies have shown that this has a crucial impact on the fRG flow by allowing for unconventional instabilities\cite{uebelacker}.
\begin{figure}[t!]
\centering
 \includegraphics[height=.55\columnwidth]{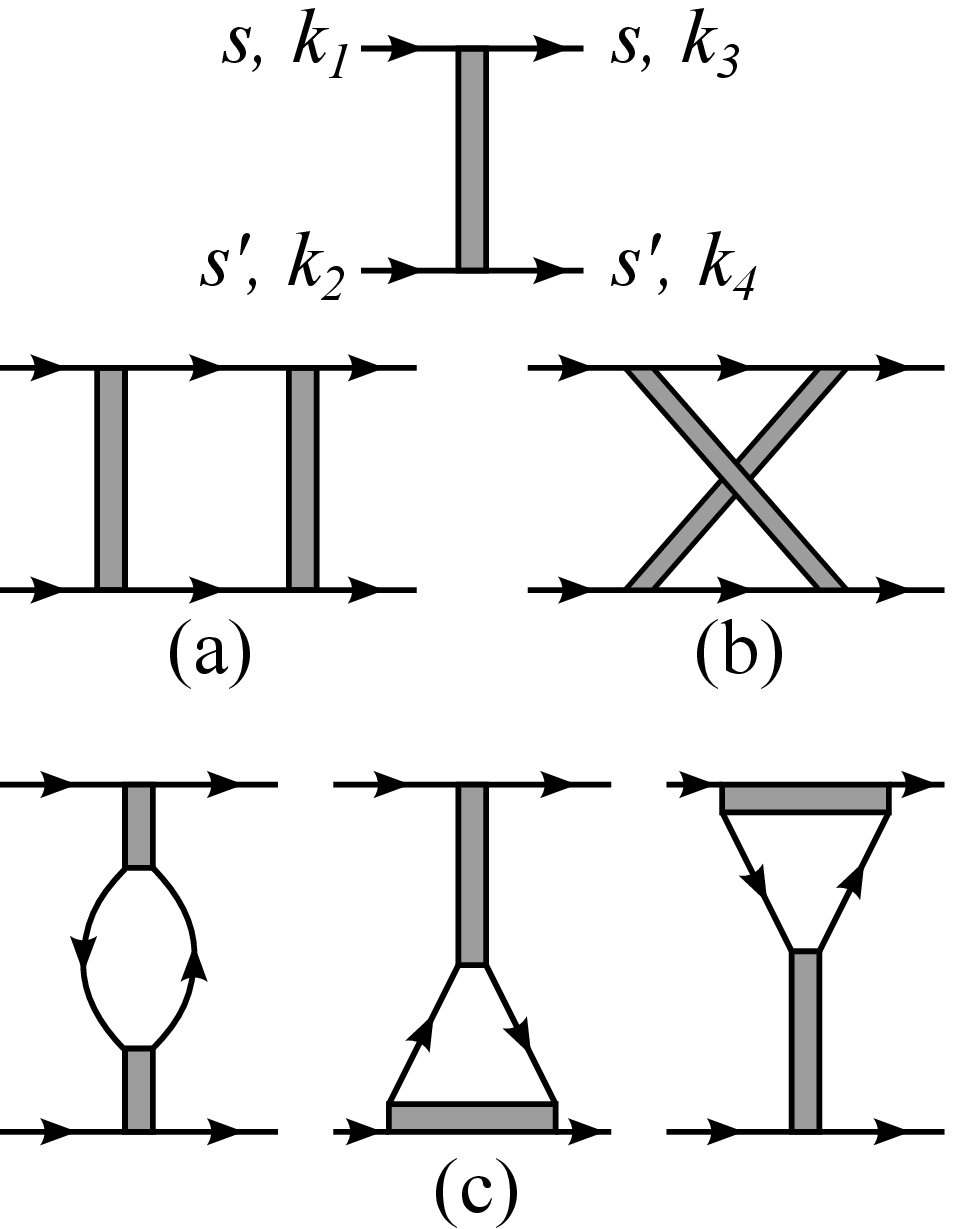}
 \includegraphics[height=.55\columnwidth]{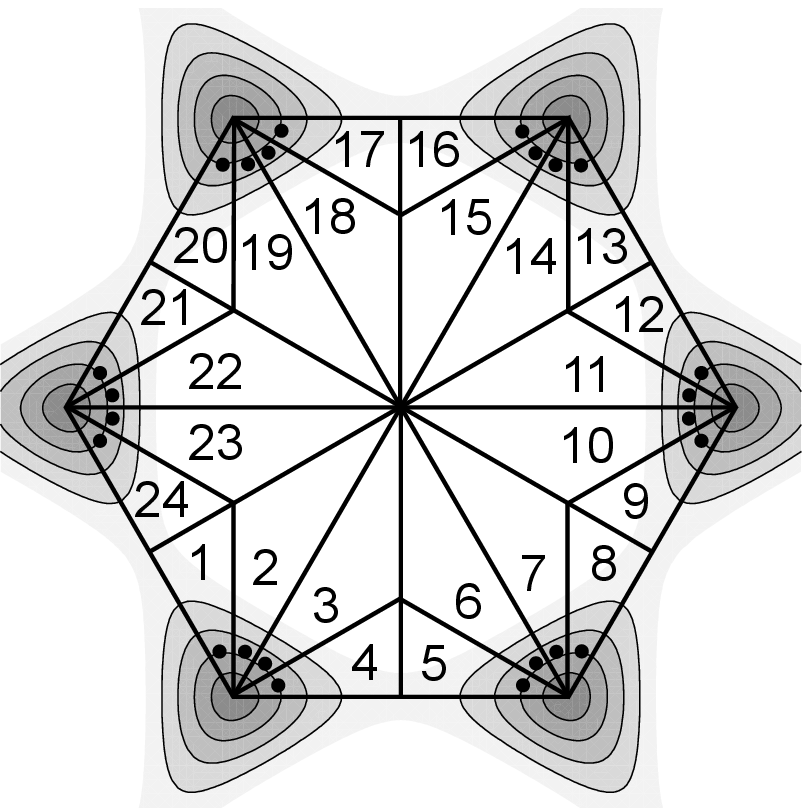}
 \caption{Left panel: Interaction vertex labeled with the spin convention (upper diagram). Below, the loop contributions to the flow of the interaction vertex including the particle-particle-diagram (a), the crossed particle-hole-diagram (b) and the direct particle-hole diagrams (c). Right panel: Patching of the Brillouin zone as explained in the text.}
\label{fig:patching}
\end{figure}

With the approximations discussed above the fRG flow for the coupling function reads
\begin{align}
\label{dgl_v}
&\frac{d}{d\Lambda}V^\Lambda ( k_1,k_2,k_3,b_4 ) =\tau^\Lambda_{PP}+\tau^\Lambda_{PH,d}+ \tau^\Lambda_{PH,cr}\, ,
\end{align}
with the particle-particle channel
\begin{align}
\label{pp}
\tau^\Lambda_{PP}&( k_1,k_2,k_3,b_4 )=- \frac{1}{V_{\text{BZ}}} \int d\vec{k} \sum_{b,b'} \Bigl[\notag\\
&V^\Lambda ( k_1,k_2,k,b' ) L^\Lambda ( k,q_{PP} ) V^\Lambda ( k,q_{PP},k_3,b_4 ) \Bigr]  \, , 
\end{align}
the direct particle-hole channel
\begin{align}
\tau&^\Lambda_{PH,d}( k_1,k_2,k_3,b_4) = -\frac{1}{V_{\text{BZ}}}  \int d\vec{k} \sum_{b,b'} \Bigl[ \notag\\
&-2V^\Lambda(k_1,k,k_3,b') L^\Lambda(k,q_{PH,d} ) V^\Lambda(q_{PH,d},k_2,k,b_4) \notag\\
& + V^\Lambda(k,k_1,k_3,b') L^\Lambda(k,q_{PH,d} ) V^\Lambda(q_{PH,d},k_2,k,b_4) \notag\\
&+ V^\Lambda(k_1,k,k_3,b')L^\Lambda(k,q_{PH,d} ) V^\Lambda(k_2,q_{PH,d},k,b_4)    \Bigr] \, , \label{phd} 
\end{align}
and the crossed particle-hole channel
\begin{align}
\label{phcr}
\tau^\Lambda_{PH,cr}&( k_1,k_2,k_3,b_4 )= - \frac{1}{V_{\text{BZ}}}  \int d\vec{k} \sum_{b,b'} \Bigl[  \notag\\ 
&V^\Lambda(k,k_2,k_3,n') L^\Lambda(k,q_{PH,cr} ) V^\Lambda(k_1,q_{PH,cr},k,b_4) \Bigr] \, , 
\end{align}
where $k=(\vec{k},b)$ collects the wave vector and the band index. As mentioned above, external frequencies are set to zero $\omega_1=\omega_2=\omega_3=\omega_4=0$.  $\vec{q}_{PP}=-\vec{k}+\vec{k}_1+\vec{k}_2$, $\vec{q}_{PH,d}=\vec{k}+\vec{k}_1-\vec{k}_3$, $\vec{q}_{PH,cr}=\vec{k}+\vec{k}_2-\vec{k}_3$ are the wavevectors of the second loop line. The band index of the second loop line is denoted with $b'$.  The frequency of the second line is fixed by frequency conservation to be $-\omega$ in the particle-particle diagram and $\omega$ in the particle-hole diagrams. $V_{\text{BZ}}$ ist the volume of the BZ. The loop kernel is given by
\begin{equation}
L^\Lambda(k,k')= \frac{d}{d\Lambda} \left[ G_0^\Lambda(k) G_0^\Lambda(k') \right] \, ,
\end{equation}
where in  our  approximation  self-energy corrections are neglected, i.e., the full propagator is identical to the free propagator.

The wavevector dependence of the interaction vertex is simplified by discretization. The BZ is divided into $N$ patches with constant wavevector dependence within one patch, so that the coupling function has to be calculated for only one representative momentum in each patch. The representative momenta for the patches are chosen to lie close to the Fermi level. The patching scheme is shown in Fig.~\ref{fig:patching}, with $N=24$. Each of the four momenta in $V^{\Lambda}(k_1,k_2,k_3,b_4)$ is additionally equipped with a band index. Momentum conservation fixes one of the four wavevectors. Altogether this results in a $6^4\times N^3$ component coupling function $V^{\Lambda}$. 

We start the fRG flow at the initial scale $\Lambda_0$ which is in our case chosen as the maximum energy of all bands. We then integrate out all modes of these bands by decreasing $\Lambda$.
In typical flows some components of the effective interaction vertex become large and diverge at a critical scale $\Lambda_c > 0$. In this work we use the scale at which the interaction vertex exceeds a value of the order of 10 times the bandwidth as an estimate for the critical scale. The precise choice of this value has only a minor effect on the extracted critical scale, as the couplings grow very fast in the vicinity of the divergence. 

The divergence is strictly speaking a (physically meaningful) artifact caused by the neglect of the self-energy in the flow. With self-energy correction a gap would open up or some other modification of the low-energy spectrum would take place, and the flow would be regularized. This is all well known from the Cooper instability in superconductors. Our analysis here tells us in which channel ordering occurs most prominently. 
The pronounced momentum structure of the vertex near the critical scale can be used to extract an effective Hamiltonian for the low-energy degrees of freedom. This is used to determine the leading order parameter of a given instability. 
Furthermore, the scale $\Lambda_c$ can be interpreted as an estimate for ordering temperatures, if ordering is allowed by the Mermin-Wagner theorem, or at least as the temperature below which the dominant correlations should be clearly observable. Furthermore, one can understand $\Lambda_c$ as energy scale for the modification of the spectrum, typically by a gap. 

In this work we study the flow at temperature $T=0$. We find flows to strong coupling with non-zero critical scales $\Lambda_c$ for all choices of non-vanishing interaction terms provided  there is a non-vanishing density of states at the Fermi level of the coupled layers.


\section{Instabilities and phase diagram}
\label{frgresults}


\subsection{ABA and ABC trilayer Hubbard model}

Let us start the description of the fRG results with the case of onsite interactions only, i.e. $U>0$, $V_1=V_2=0$. We limit the study to the charge-neutrality point, i.e. with Fermi points at $K$ and $K'$ in the Brillouin zone.
%
%
\subsubsection{Simplified band structures}
\begin{figure}[t!]
\centering
 \includegraphics[height=.27\columnwidth]{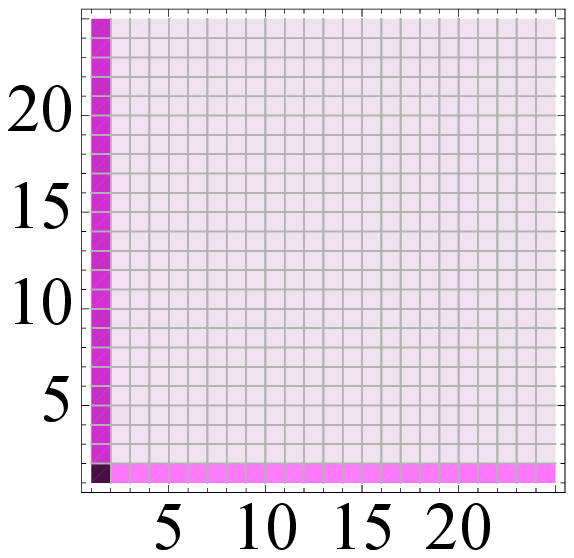}
  \includegraphics[height=.27\columnwidth]{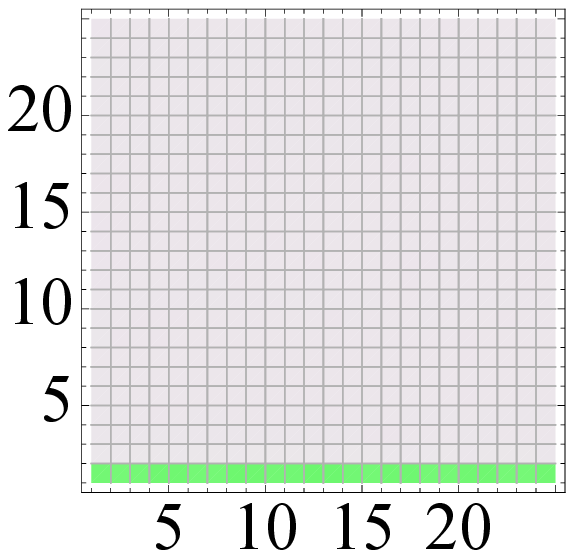}
   \includegraphics[height=.27\columnwidth]{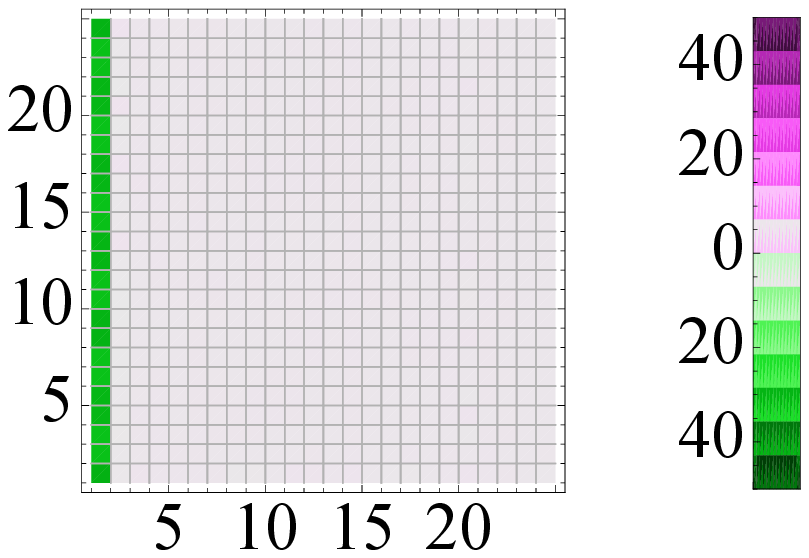}
 \caption{Effective interaction vertex near the critical scale in the AF regime in units of $t$. Left Panel: Orbital combinations with $o_1=o_2=o_3=o_4$. The numbers on the axis specify the number of the patch as shown in Fig.~\ref{fig:patching}. On the horizontal axis the wavevector $k_1$ can be read off and on the vertical axis we enumerate $k_2$. $k_3$ is fixed on the first patch, $k_4$ then follows from momentum conservation. Here, we see that the sharp vertical structure ($k_1=k_3$) comes with the double magnitude as the horizontal structure ($k_2=k_3$). Middle Panel: Effective vertex function for the orbital combination, where $o_1=o_3, o_2=o_4\neq o_1$ and if $o_1 \in \{a_1,a_2,a_3\}$ then $o_2 \in \{b_1,b_2,b_3\}$. Right panel: Effective vertex function for the orbital combination, where $o_1=o_4, o_2=o_3\neq o_1$ and if $o_1 \in \{a_1,a_2,a_3\}$ then $o_2 \in \{b_1,b_2,b_3\}$. If in the second and third plot both $o_1, o_2 \in \{a_1,a_2,a_3\}$ or $\in \{b_1,b_2,b_3\}$ the sign of the vertices changes.}
\label{fig:ABCSDW}
\end{figure}
First, we want to investigate the simpler band structures when all remote hopping terms are neglected and only $t\neq 0$ and $t_\perp\neq0$. Then, running the fRG flow with pure onsite interaction $U\neq 0$ in the ABC and ABA stacked trilayer, we observe an instability toward an antiferromagnetic spin density wave (AF-SDW) with a typical signature of the interaction vertex near the instability as shown in Fig.~\ref{fig:ABCSDW}. The leading part of effective interaction corresponding to the clearly discernible sharp structures in wave vector space reads in this case
\begin{equation}
H_{\mathrm{AF}} =-\frac{1}{\mathcal{N}} \sum_{o,o^\prime} V_{oo^\prime}\epsilon_o \epsilon_{o^\prime}\vec{S}_{\vec{q}=0}^{o}\cdot \vec{S}_{\vec{q}=0}^{o^\prime}
\end{equation}
with $V_{oo'}>0$ and $\vec{S}_{\vec{q}=0}^{o}=\frac{1}{2} \sum_{\vec{k},s,s'}\vec{\sigma}_{ss^\prime} c^{\dagger}_{o,s,\vec{k}} c_{o,s^\prime,\vec{k}}$.  The fact that the above Hamiltonian only contains the ${\vec{q}=0}$ component means that the effective interaction has become infinitely-ranged\cite{Honerkamp2008}. The  factors $\epsilon_o$ depend on the orbital, $\epsilon_o=+1$ for $o\in \{a_1,a_2,a_3\}$ and $\epsilon_o=-1$ for $o \in \{b_1,b_2,b_3\}$. This sign structure implements the staggering of the interaction within the unit cell appropriate for antiferromagnetic interactions.  Note that this parametrization holds in both cases, for the ABC as well as for the ABA stacking.

A mean-field decoupling of $H_{\mathrm{AF}}$ results in an AF spin alignment in each layer where a net spin (e.g. `up') moment is located on the A$_1$-, A$_2$- and A$_3$ sublattices, and an opposite net spin (`down') moment on the B$_1$-, B$_2$- and B$_3$-sublattices. 

\begin{figure}[t!]
\centering
 \includegraphics[width=1.0\columnwidth]{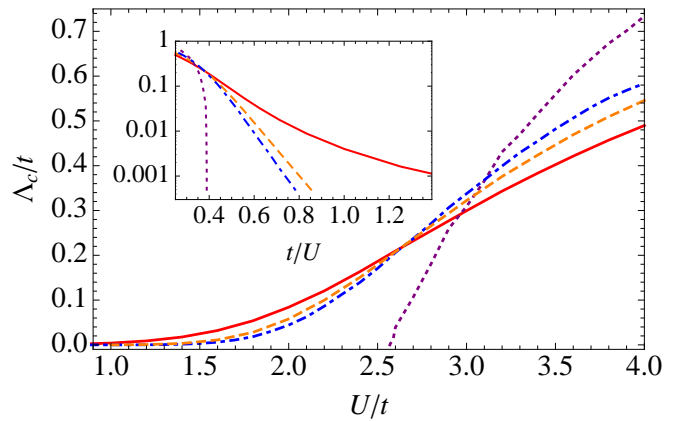}
 \caption{fRG critical scale for the singlelayer (dotted purple), the AB bilayer (dot dashed blue), the ABA trilayer (dashed orange) and the ABC trilayer (solid red). These results have been obtained with $t_\perp=t$ and all higher hopping terms set to zero. The inset shows the same data in a semilog plot over $t/U$.}
\label{fig:hubbardU}
\end{figure}

The critical scale $\Lambda_c$ as a function of the onsite interaction $U$ for ABC and ABA honeycomb trilayers with model hopping parameters $t=t_\perp$ is shown in Fig.~\ref{fig:hubbardU} together with the critical scales of single- and bilayer honeycomb lattices with the same hopping parameters. This choice of band parameters takes us beyond the regime of realistic parameters for TLG, but pronounces the characteristic features of the ABC and the ABA stacking close to the $K, K^\prime$ points and therefore allows to study the differences of the various honeycomb stacks more explicitly. Furthermore, it allows to compare to recent QMC results for the bilayer system\cite{lang2012}. We add a systematic study of the dependence on $t_\perp$ below.
Most importantly, we observe that in the ABC trilayer, the critical scale decreases more slowly as compared to AB bilayer when the onsite interaction is decreased. While in the case of ABA trilayer and AB bilayer with quadratic band crossing points the functional dependence of $\Lambda_c(U)$ can be fitted by an exponential decay $\sim \exp(-\alpha/U)$ (cf. inset of Fig.~ \ref{fig:hubbardU}), this does not hold for the ABC trilayer case. Instead, at small $U$, based on the density of states $\sim \epsilon^{-1/3}$, one would naively expect a behavior $\Lambda_c \sim U^3$. This is however not reproduced by our data, presumably due to the influence of the high energy sector, i.e. additional bands. We expect that the leading $\Lambda_c \sim U^3$ dependence might be recovered at smaller $U$ and thus smaller $\Lambda$, which is numerically hard to access due to the rapidly decreasing critical scale.
From this analysis we conclude that in ABC trilayers  an interaction-induced gap, which we expect to be of the order of the critical scale $\Lambda_c$, may be considerably larger than in the other structures. This also implies a more stable correlated ground state. All this seems compatible with recent experiments \cite{bao2011}.

%
%
\subsubsection{Inclusion of remote hoppings in ABA trilayers}
When the remote hoppings $\gamma_2,\gamma_3,\gamma_4,\gamma_5$ and the onsite energy $\delta$ in ABA trilayers are taken into account, the band dispersion is deformed considerably and the particle-hole nesting of the band structure is destroyed\cite{fzhang2012,mccann2011}, see Fig.~\ref{fig:trilayerdisp}. In this case, one should expect that  in the weakly interacting limit the tendency toward instabilities will be strongly reduced. However, it is interesting to study what happens in the case of intermediate and larger onsite interactions. 

An accurate treatment of the dispersion with remote hoppings at lowest scales would require the implementation of a new patching scheme to resolve the vicinity of the $K$ and $K'$ points, i.e. the fact that the Fermi surface is now not restricted to  a single point. This is beyond the scope of this work. Therefore we now stop the flow at the energy scale $\Lambda^*$ at which the bands become non-monotonous. This procedure is routinely done in parquet- and $g$-ology studies of imperfectly nested bands\cite{dwcutoff}.  Of course it leaves the scales below $\Lambda^*$ unintegrated, but as the dispersion at these lowest scales is not nested and partially gapped, we do not expect significant contributions of these modes to the flow. Hence we expect that the so-obtained estimate for critical scales and $U_c$ is already quite good.
For the hopping parameters given in Refs.~\onlinecite{fzhang2012,mccann2011} the energy scale at which the dispersion is not monotonous any more is given by $\Lambda^* \sim 10\,\mathrm{meV}\sim 0.004t$. 

\begin{figure}[t!]
\centering
 \includegraphics[width=.9\columnwidth]{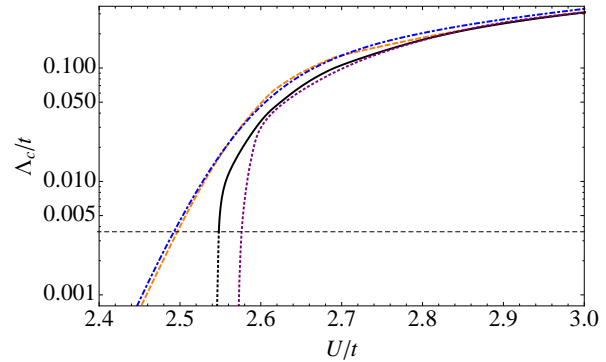}
 \caption{Critical scale of ABA-trilayer including remote hoppings (denoted ABA$^\ast$, solid black line), with the ratios of the hopping parameters corresponding to the realistic values\cite{fzhang2012,mccann2011},  and with pure onsite repulsion $U$ as interaction. For comparison, we also show the critical scales of ABA-trilayer without remote hoppings (dashed orange), AB bilayer (dot dashed blue), both with the choice $t_\perp=0.1t$, and the single-layer system (dotted purple). The dashed horizontal line visualizes the energy scale $\Lambda^\ast$ where the bands become non-monotonous due to remote hoppings.}
\label{fig:ABAast}
\end{figure}

However, it is important to notice that the remote hoppings already change the dispersion far above the scale $\Lambda^*$. This explains why for the ABA stacking (here from now on called ABA$^*$ stacking when the remote hopping are included), the effect of the remote hoppings is quite drastic. Studying again the case of onsite interactions $U$ only, 
we observe clearly diverging susceptibilities only for onsite interactions $U \gtrsim 2.6t$ at critical scales well above $0.004t$. For smaller interactions, the couplings grow very slowly and no divergences at finite scales above $\Lambda^*$ can be identified. 
In Fig.~\ref{fig:ABAast} we show the fRG results for the critical scales for the ABA$^*$ trilayer with remote hopping terms. For comparison, we also show the curves for the single- and bilayer system. As the similarity to the single layer is strong, this analysis suggests a critical onsite interaction $U_c\approx 2.6t$ above which a many-body instability can occur in the ABA$^*$ stacking with remote hopping included. We would like to add that the fRG in the present approximation has the tendency to overestimate critical scales. Therefore we would expect  the true $U_c$ to be slightly larger. For instance, in the single layer system for onsite interactions only, QMC gives  $U_{c,\mathrm{QMC}}\approx 3.4t$ for the opening of a single-particle gap\cite{meng}, while in the fRG we find $U_c \approx 2.6t$ as well (see Fig.~\ref{fig:hubbardU}).  

The experimental study of Ref.~\onlinecite{bao2011} does not find a gap for ABA($^*$) trilayer graphene. This is compatible with our findings, given the interactions in real material are weaker than this critical value. This is to be expected for consistency,  as the critical interaction strength of the single layer and the ABA$^* $-trilayer are roughly the same, and the single-layer remains  semi-metallic, too.
Even if they are slightly above the threshold, the expected transport gaps would be small and very hard to measure. 
Therefore, while a precise quantitative picture cannot be obtained with our approximate fRG method and here for onsite interactions only, on a qualitative level we reach consistent conclusions. It would be interesting to resolve better the critical region close to $U_{c,\mathrm{TL}}$ for the ABA$^*$ stacked trilayer with remote hoppings and analyze the onset of instabilities and their nature. However, this would require a different implementation of our patching scheme which we leave out for future work. In the remainder of this work, we will therefore concentrate on a more thorough study of the ABC trilayer model.
%
%

\subsection{ABC trilayer: Instabilities with non-local interactions}

From the results for pure onsite interactions and also inspired by recent experiments, we conclude that while ABC trilayers strongly support the formation of many-body states, the ABA$^*$ trilayer including remote hoppings most probably does not. We take this as a motivation to study further the instabilities of ABC trilayers for wider range of non-local interaction parameters, which brings us closer to the real materials.
\emph{Ab initio} values for the strength of the density-density interactions up to the third nearest neighbor were listed for single-layer graphene and graphite in Ref.~\onlinecite{wehling2011}. Most likely, one can safely interpolate the parameters for the bi- and trilayer case from this data.
 
Running the fRG for extended interactions, we find a number of different phases as previously described for the bilayer case in Ref.~\onlinecite{scherer2012}, namely a charge density wave (CDW), a quantum spin Hall state (QSH) and charge density wave with non-zero momentum transfer (CDW$_3$) alongside the AF-SDW. 
For the investigation of the phase diagram we take into account non-local interaction contributions, namely the nearest-neighbor in-plane interaction $V_1$ and the second-nearest neighbour in-plane interaction $V_2$. More remote interaction contributions are neglected. In the bilayer case\cite{scherer2012} we checked explicitly that a third-nearest neighbor repulsion $V_3$ does not change the picture. Also, we do not consider  interlayer interactions. The ab-initio calculations in Ref.~\onlinecite{wehling2011} showed that the nearest interlayer interactions in graphite (as well as in layered graphene) are of the order of the $V_3$-term\cite{wehlingprivate}. 

\begin{figure}[t!]
\centering
\includegraphics[height=.27\columnwidth]{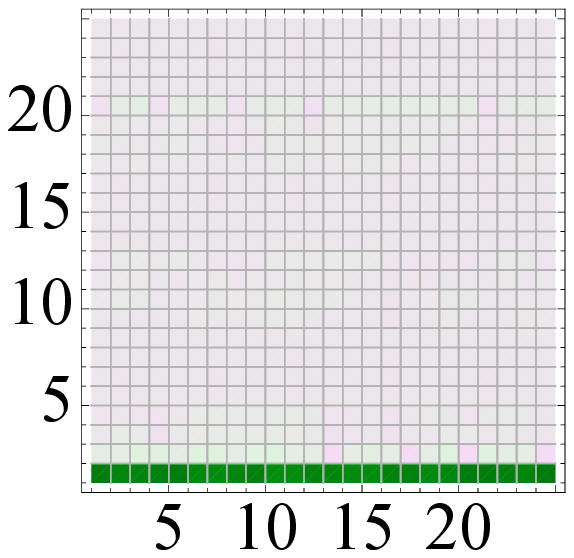}
  \includegraphics[height=.27\columnwidth]{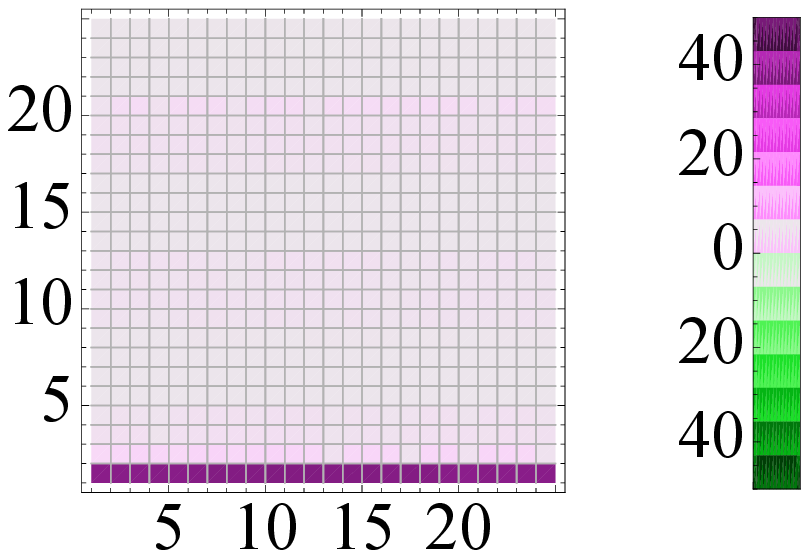} 
  \caption{Effective interaction vertex near the critical scale in the CDW regime for $U=0$, $V_1=0.5t$ and $V_2=0$  in units of $t$. Left Panel: Orbital combinations with $o_1=o_2=o_3=o_4$. The numbers on the axis specify the number of the patch as shown in Fig.~\ref{fig:patching}. On the horizontal axis the wavevector $k_1$ can be read off and on the vertical axis we enumerate $k_2$. $k_3$ is fixed on the first patch, $k_4$ then follows from momentum conservation. Left Panel: Effective vertex function for the orbital combination, where $o_1=o_3, o_2=o_4\neq o_1$ and if $o_1 \in \{a_1,a_2,a_3\}$ then $o_2 \in \{b_1,b_2,b_3\}$.  If in the right plot both $o_1, o_2 \in \{a_1,a_2,a_3\}$ or $\in \{b_1,b_2,b_3\}$ the sign of the vertices changes.}
\label{fig:CDW}
\end{figure}

The two leading non-local interaction terms $V_1$ and $V_2$ trigger the appearance of qualitatively different instabilities. For dominating $V_1$ we find an instability toward a charge density wave with a momentum signature of the effective interaction as shown in Fig.~\ref{fig:CDW}. This momentum structure can be written down as an effective interaction Hamiltonian of the form,
\begin{equation}
H_{\mathrm{CDW}}=-\frac{1}{\mathcal{N}} \sum_{o, o'}V_{oo'} \epsilon_o  \epsilon_{o'} N^{o}N^{o'}       
\end{equation}
with $V_{oo'}>0$ and $N^{o}=\sum_{\vec{k},s}c^{\dagger}_{o,s,\vec{k}} c_{o,s,\vec{k}}$. This sign structure supports an enhanced occupancy of the $A_i$ sublattices as compared to the $B_i$ sublattices
 or vice versa. Furthermore, a mean-field decoupling of this effective interaction gives a gap in the single-particle spectrum.

A dominating $V_2$ yields an instability whose dominant interaction terms can be cast into an effective Hamiltonian of the following type,
\begin{equation}
H_{\mathrm{QSH}}=- \frac{1}{\mathcal{N}} \sum_{o, o'}V_{oo'} \epsilon_o \epsilon_{o'}\vec{S}^{o}_f\cdot\vec{S}^{o'}_f
\end{equation}
with $V_{oo'}>0$ and $\vec{S}^{o}_f=\frac{1}{2}  \sum_{\vec{k},s,s'}f_{\vec{k}}\vec{\sigma}_{ss^\prime} c^{\dagger}_{o,s,\vec{k}} c_{o,s^\prime,\vec{k}}$ including a $f$-wave form factor $f_{\vec{k}}=\sin(\sqrt{3}a k_x)-2 \sin(\frac{\sqrt{3}a k_x}{2}) \cos(\frac{3a k_y}{2})$. 
This effective Hamiltonian can be decoupled in a purely imaginary `Kane-Mele' order parameter. This type of instability represents a many-body path to the quantum spin Hall state, where the mass term due to spin-orbit interaction in the original Kane-Mele proposal\cite{kanemele2005} is now provided by an interaction-induced mean-field. In the single-layer\cite{HonerkampRaghu2008} and bilayer\cite{scherer2012} honeycomb models, this instability was found in the same corner of interaction parameter space. For an odd number of layers, the mean-field Kane-Mele-ordered state will give rise to an odd number of helical edge modes, and thus represent a two-dimensional interaction-driven topological insulator with protected edge modes.
\begin{figure}[t!]
\centering
 \includegraphics[height=.27\columnwidth]{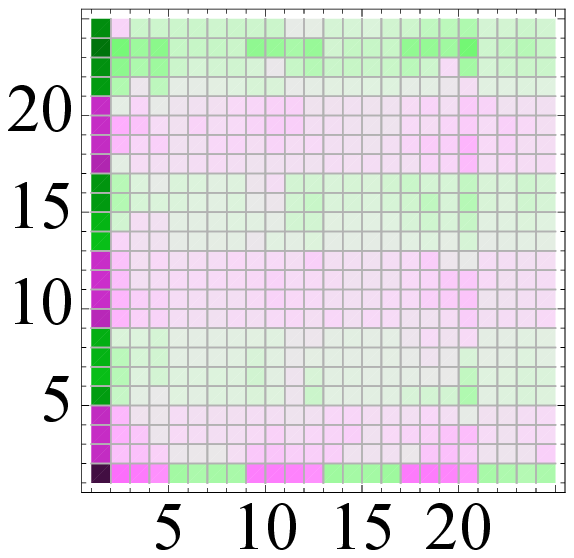}
  \includegraphics[height=.27\columnwidth]{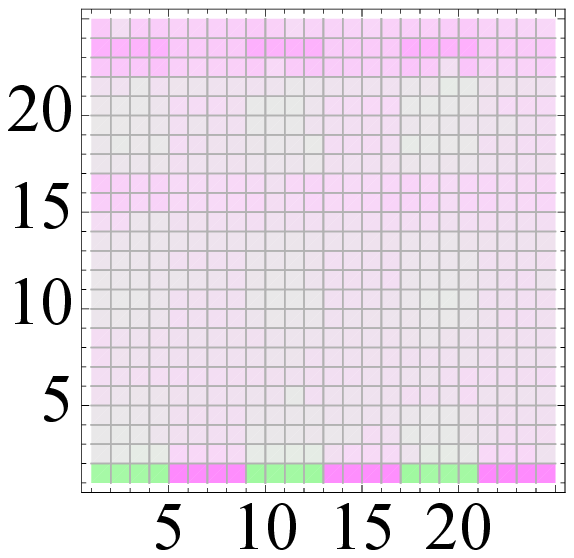}
   \includegraphics[height=.27\columnwidth]{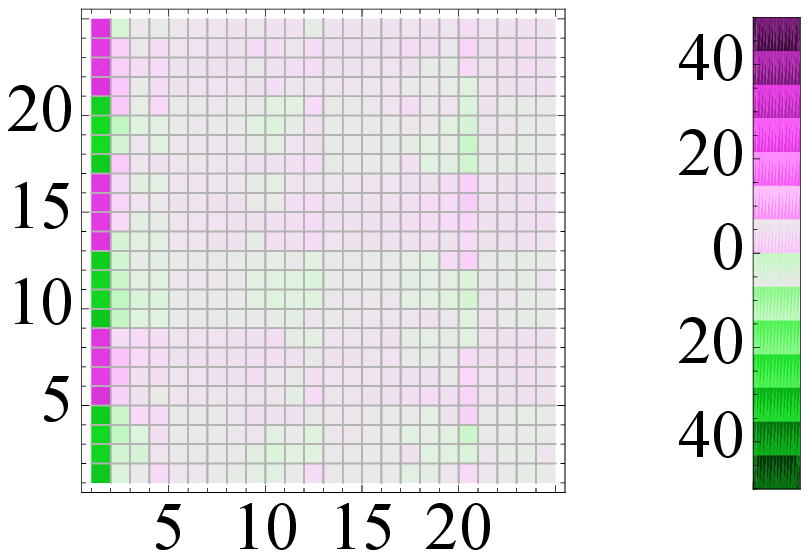}
 \caption{Effective interaction vertex near the critical scale in the QSH regime for $U=0$, $V_1=0$ and $V_2=1.5t$ in units of $t$. Left Panel: Orbital combinations with $o_1=o_2=o_3=o_4$. The numbers on the axis specify the number of the patch as shown in Fig.~\ref{fig:patching}. On the horizontal axis the wavevector $k_1$ can be read off and on the vertical axis we enumerate $k_2$. $k_3$ is fixed on the first patch, $k_4$ then follows from momentum conservation. Here, we see that the sharp vertical structure ($k_1=k_3$) comes with the double magnitude as the horizontal structure ($k_2=k_3$). Middle Panel: Effective vertex function for the orbital combination, where $o_1=o_3, o_2=o_4\neq o_1$ and if $o_1 \in \{a_1,a_2,a_3\}$ then $o_2 \in \{b_1,b_2,b_3\}$. Right panel: Effective vertex function for the orbital combination, where $o_1=o_4, o_2=o_3\neq o_1$ and if $o_1 \in \{a_1,a_2,a_3\}$ then $o_2 \in \{b_1,b_2,b_3\}$. If in the second and third plot both $o_1, o_2 \in 
 \{a_1,a_2,a_3\}$ or $\in \{b_1,b_2,b_3\}$ the sign of the vertices changes .}
\label{fig:ABCQSH}
\end{figure}

\begin{figure}[t!]
\centering
\includegraphics[height=.27\columnwidth]{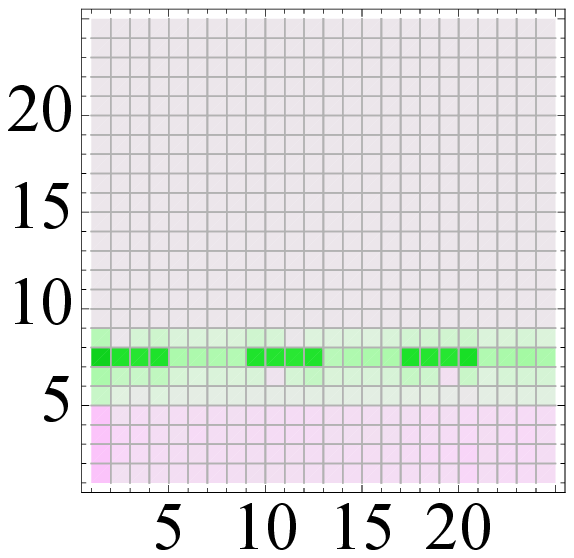}
  \includegraphics[height=.27\columnwidth]{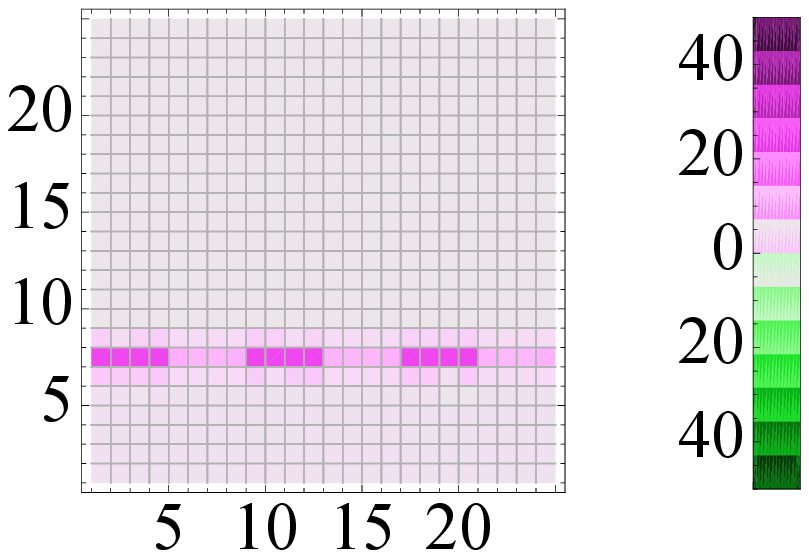} 
  \caption{Effective interaction vertex near the critical scale in the CDW$_3$ regime for $U=0$, $V_1=0$ and $V_2=0.5t$ in units of $t$. Left Panel: Orbital combinations with $o_1=o_2=o_3=o_4$. The numbers on the axis specify the number of the patch as shown in Fig.~\ref{fig:patching}. On the horizontal axis the wavevector $k_1$ can be read off and on the vertical axis we enumerate $k_2$. $k_3$ is fixed on the first patch, $k_4$ then follows from momentum conservation.  Here, we see that the order parameter has non vanishing momentum transfer as the sharp feature is not at $k_1=k_3$. Right Panel: Effective vertex function for the orbital combination, where $o_1=o_3, o_2=o_4\neq o_1$ and if $o_1 \in \{a_1,a_2,a_3\}$ then $o_2 \in \{b_1,b_2,b_3\}$. If in the right plot both $o_1, o_2 \in \{a_1,a_2,a_3\}$ or $\in \{b_1,b_2,b_3\}$ the sign of the vertices changes.}
\label{fig:CDW3}
\end{figure}

Finally, in the niche of the parameter space for smaller $U$, we also recover the CDW$_3$ phase, that we already found in the bilayer system\cite{scherer2012},
\begin{equation}
H_{\mathrm{CDW_3}}=- \frac{1}{\mathcal{N}} \sum_{o,o'}V_{oo'} \epsilon_o \epsilon_{o'} \big(N^{o  }_{\vec{Q}} N^{o'}_{-\vec{Q}}+N^{o}_{-\vec{Q}} N^{o'}_{\vec{Q}}\big)
\label{hcdw3} 
\end{equation}
with $N^{o}_{\vec{Q}}=\sum_{\vec{k},s}c^{\dagger}_{o,s,\vec{k}+\vec{Q}} c_{o,s,\vec{k}}$. See Fig.~\ref{fig:CDW3} for the characteristic momentum structure
of the effective interaction. The order parameter due to the symmetry breaking $\langle N^{o}_{\vec{Q}} \rangle \not= 0$ is in complete analogy to the one in the honeycomb bilayer\cite{scherer2012}, except for the adapted definition of the $\epsilon_o$. Within one layer, this order forms three inequivalent sites with different charge densities. The sign structure of the order parameter on different layers and sublattices is determined by the $\epsilon_o$-factors, so as to lower the energy contribution from (\ref{hcdw3}). This leaves the total phase of the order parameter undetermined. Depending on this phase, the quadratic mean-field Hamiltonian for a single layer either exhibits a gapless spectrum with Dirac points shifted away from the $K$, $K'$ points or a fully gapped state. Which case represents the mean-field ground state as function of the interaction parameters has yet to be determined variationally.
 

\subsection{ABC trilayer phase diagram from fRG}

\begin{figure}[t!]
\centering
 \includegraphics[width=1.0\columnwidth]{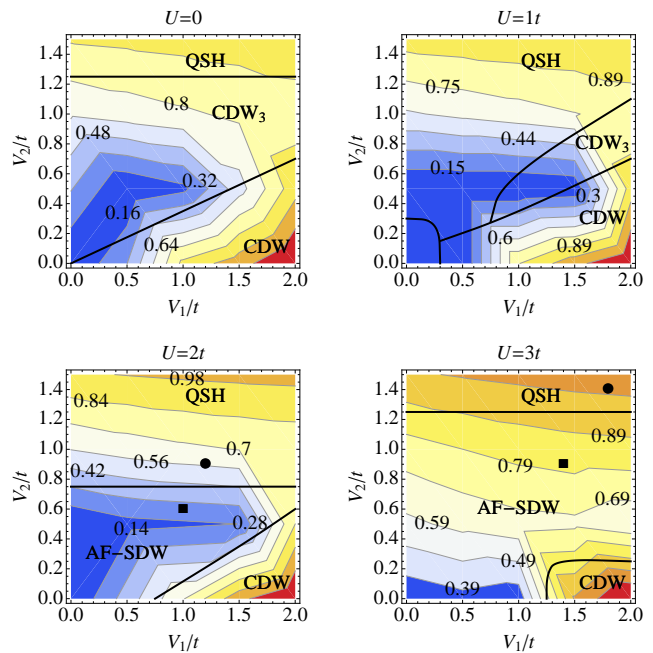}
 \caption{Tentative fRG phase diagram of the ABC trilayer with $t_\perp=0.1 t$. The black lines are guides to the eye and seperate the different regimes. The contourplot encodes the critical scale at which the vertices diverge. The rescaled ab-initio interaction parameters for graphene (circles) and graphite (squares) of Ref.~\onlinecite{wehling2011} are shown in the lower plots.}
\label{fig:phasediagABC}
\end{figure}

For a systematic investigation of the ABC trilayer phase diagram, we scan a range of interaction parameters $U$, $V_1$ and $V_2$, whose ab-initio values are listed in Ref.~\onlinecite{wehling2011}. As we expect the fRG to overestimate the critical scales, we take these ab-initio parameters as upper bounds for the range our phase diagrams. In Fig.~\ref{fig:phasediagABC}, we show the fRG phase diagram obtained by identifying the leading tendencies in the effective interactions near the instability with an underlying contour plot of the critical scale $\Lambda_c$ in units of $t$.

We also mark the ab-initio values in the lower plots of Fig.~\ref{fig:phasediagABC}, obtained by taking the values from Ref.~\onlinecite{wehling2011} and scaling $\{ U, V_1,V_2\} \to \alpha \{ U, V_1,V_2\}$ so as to hit the values $U=2t$ and $U=3t$. In both cases, these choices place the system near the phase boundary between QSH and AF-SDW instability. For the more long-ranged single-layer graphene interactions, one finds a QSH state, while for the slightly shorter ranged graphite parameters, one gets a AF-SDW state. Hence, which order occurs might be a delicate issue that is decided by details. In our approximation, the critical scales interpolate smoothly across the phase borders, indicating a weaker competition between the different tendencies.
Note that due to this borderline situation there is no true necessity for different layered graphene systems, e.g. with different environments, to exhibit the same ground state, and even the energy scales or gaps could come out similarly despite different states might be selected. Hence, regarding the leading instability, the situation in the ABC trilayer is very similar to the one found previously in the Bernal-stacked bilayer system.

As mentioned before, we also find a rather exotic density wave phase CDW$_3$ with a tripling of the unit cell within the layers. This state is subject to further investigation. However it occurs only at quite unrealistic corners of the parameter space with dominant non-local terms. Hence, we do not discuss it further here.
 

 \subsection{Energy scales of the ordering phenomena}
 \label{energyscales}
In our previous study of the bilayer system\cite{scherer2012},  we encountered a problem of the current model studies that in principal also affects the present work. However, we will now also offer an explanation of what happens and how one should read the data in order to get reasonable agreement with and a more quantitative picture of the experiments.
 
As shown in this work and also in the previous paper on the bilayer system\cite{scherer2012}, the simple models employed by virtually all many-body approaches to interaction effects in few-layer graphene can produce very large critical scales. 
This can be already seen in Fig.~\ref{fig:phasediagABC}, where we indicate the values of the critical scales in units of the hopping $t\approx 3\,$eV. These scales are huge for most choices of the interaction parameters. This is not surprising.
In our and other theoretical approaches, the large scales are simply caused by the perfect particle-hole nesting of the band structure. Furthermore, from the comparison with QMC calculations in the case of pure onsite interactions\cite{lang2012} we know that the overestimate of the fRG is certainly not severe and, expressed conservatively, is less than an order of magnitude in the regime where also the QMC finds robust ordering. As the critical scale is an estimate for the energy scale of the spectral restructuring or gap opening in the ordered phase, a high critical scale would correspond to large energy gaps.
 If we took the ab-initio parameters of Ref.~\onlinecite{wehling2011} literally, 
the theoretical gap estimates would exceed the experimentally observed gap scales $\sim 1-10\,$meV by orders of magnitude. 
\begin{figure}[t!]
\centering
  \includegraphics[width=1.0\columnwidth]{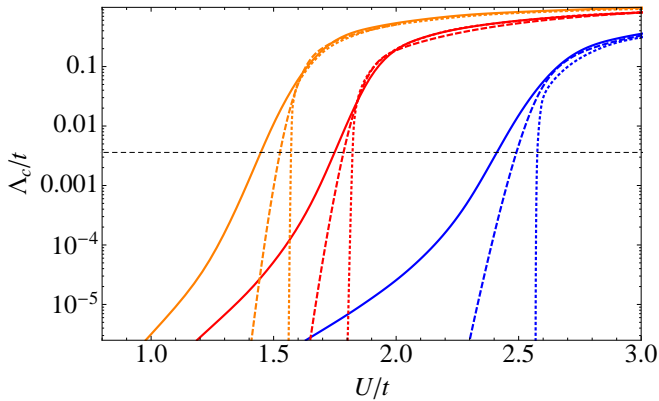}
 \caption{FRG critical scale for ABC trilayer for the pure Hubbard model (blue solid) and rescaled \emph{ab initio} parameters for graphite (red solid)  and single-layer graphene (orange solid) as described in the text with $t_\perp=0.1 t$. For comparison we also show the corresponding critical scale for the AB bilayer structure (dashed) and the singlelayer structure (dotted). For the Hubbard (blue curves) and the rescaled graphite parameters (red curves) we find the system to be in the AF-SDW phase, for the rescaled graphene parameters the system is in the QSH state (orange curves). The dashed horizontal line marks 10 meV, which is the order of magnitude, where the topology of the band structure in ABC trilayer graphene changes and remote hoppings become important.}
\label{fig:rescaled}
\end{figure}

\begin{figure}[t!]
\centering
 \includegraphics[width=1.0\columnwidth]{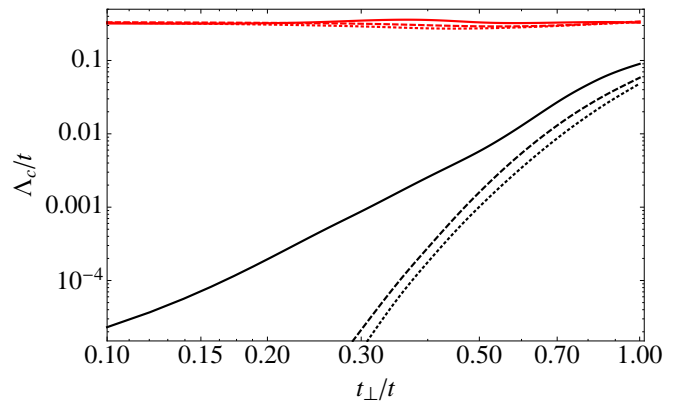}
 \caption{FRG critical scale for ABC trilayer (solid lines), ABA trilayer without remote hoppings (dashed lines) and AB bilayer (dotted lines) with fixed onsite interaction and variable $t_\perp$. The red lines show the results for $U=3t > U_{c,\mathrm{SL}}$ and no dependence of the critical scale on the interlayer hopping $t_\perp$. For the case $U=2t<U_{c,\mathrm{SL}}$ (black lines) we observe a strong dependence of the critical scale on $t_\perp$.}
\label{fig:vartp}
\end{figure}

Let us now analyze the systematics of these critical scales a bit more deeply.
In Fig.~\ref{fig:rescaled}, we show the critical scales obtained from fRG for the single-, bi- and trilayer graphene for three cases, namely for onsite interaction $U$ only, and for nonlocal interactions with the cRPA parameters $V_1$ and $V_2$ for graphite and graphene with repulsions up to the second nearest neighbor, for realistic interlayer hopping $t_\perp = 0.1t$. The curves show the dependence on the overall magnitude of the onsite interaction, where the ratio between the local $U$ and the non-local interaction parameters is kept fixed.
Obviously there are two regimes: a high-scale regime with large critical scales that do not depend too strongly on the interaction strength and a low-scale regime with a very strong dependence on the interaction. For the single layer the second regime is very narrow and basically only contain the minimal critical interaction strength $U_c (V_1,V_2)$ below which the semi-metal is stable. Also for bi- and trilayer, the high-scale regime starts above the single-layer $U_c (V_1,V_2)$.
Next let us consider the dependence of the critical scale $\Lambda_c$ on the perpendicular hopping $t_\perp$ in the ABC trilayer. Whereas the nature of the ground state qualitatively remains the same for all choices of $t_\perp \neq 0$ the absolute value of this parameter has an impact on the size of the critical scale and shows different behaviors on the two different sides of the critical onsite interaction of single-layer graphene $U_{c,\mathrm{SL}}$, see Fig.~\ref{fig:vartp}, i.e. whether we are in the high-scale or in the low-scale regime. For large interactions $U > U_{c,\mathrm{SL}}$ the size of $t_\perp$ is of no importance for the critical scale $\Lambda_c$. This changes for smaller $U$. Here, the smaller $t_\perp$ is, the stronger is the $\Lambda_c$-variation with the interaction strength.
The comparison with QMC in Ref.~\onlinecite{lang2012} was done at larger $t_\perp = t$ and $U \ge 2.8t$, where the scales do not vary that strongly. 
For band structure parameters with $t_\perp \lesssim 0.1t$ and small interactions $U \lesssim U_{c,\mathrm{SL}}$, we also observe flows to strong coupling for the ABC and the ABA trilayer system, however, the critical scales turn out to be very small, an effect that reflects the behavior of the single-layer system, where no instabilities occur for $U < U_{c,\mathrm{SL}}$. While this does not constitute a difficulty for the fRG method \emph{per se} it makes the numerical evaluation very tedious and renders the comparison with QMC impossible. 

We now argue that in order to account for the approximations made in the fRG scheme, mainly the neglect of self-energy corrections, and in order to obtain a realistic picture, we have to reduce the cRPA parameters with $U \sim 3t$ by hand. The reduction factor is chosen so as to obtain the experimentally verified semi-metallic solution for the single-layer case, i.e. roughly to $U \sim 1.5t$ (if we take the single-layer cRPA values) with an analogous rescaling of the non-local couplings by the same factor. This shift on the interaction axis now takes us from the high scale regime into the regime of strongly varying scales, cf. Fig. \ref{fig:rescaled}. Now the critical scales for bi- and trilayer end up in the range $\lesssim 0.01t \sim 30\,$meV which is much closer to the experimental values for gap sizes and already in the correct order of magnitude. 

In the experiments on bi- and trilayer graphene of these types, the energy gap in the trilayer\cite{bao2011} came out a factor 2-3 higher than in the bilayer. In the fRG the ratio between the critical scales for these two systems  depends on the parameter values for the interactions. However for intercations where single-layer graphene does not undergo a phase transition, the larger density of states near the Fermi level of the trilayer case leads to a larger critical scale in theory as well, see Fig.~\ref{fig:rescaled}. Hence the different energy gaps in bi- and trilayers are qualitatively captured by this theory. Remarkably, with this choice of interaction parameters, the instabilities in ABC trilayer occur on an energy scale where remote hoppings start to be important ($\sim 10\,$meV, see Fig.~\ref{fig:rescaled}).


\section{Discussion}
\label{discussion}
We have performed extensive fRG calculations on honeycomb trilayer systems with different stacking orders, as model systems for trilayer graphene. In doing this we have used as far as possible the available input parameters from ab-initio calculations. Moreover, we have taken into account the full 6-band band structure obtained within the model with one effective $p_z$-type orbital per carbon site. The fRG approach we have used is certainly not exact, but it goes far beyond mean-field studies, random-phase approximation approaches and other perturbative calculations performed for these systems. Furthermore, in the bilayer case with onsite interactions only, the qualitative and quantitative information of the fRG compares quite favorably with the results of QMC calculations\cite{lang2012}. For non-local interactions, QMC encounters severe sign problems, and there is no possibility for benchmarks, but there is also no clear reason why the fRG method should get worse then. Hence, for the given theoretical model the fRG results should be rather robust both qualitatively and also quantitatively (with the adaptations discussed is Sec. \ref{energyscales} and mentioned below) at least as an order-of-magnitude estimate. 

First let us mention the main results of the present trilayer study  before we get to the connections with graphene systems and experiments. Comparing the different trilayer stackings with the Bernal-stacked bilayer and the monolayer results, we could identify the ABC trilayer as the system that is most prone toward instabilities, with larger energy scales for ordering than the AB-bilayer. 
The ABA trilayer without the  additional interlayer `remote' hoppings is comparable in its critical scales to the bilayer, but we showed the remote hopping terms with the suggested realistic  parameter values (called ABA$^*$ structure here) are likely to remove the instability at least for smaller, possibly realistic, interaction strengths. Interestingly, for the Hubbard onsite interaction case, the minimal $U$-value for obtaining a gapped ground state in the ABA$^*$-structure is close to the one for the single layer. Taking the current experimental knowledge for bi- and trilayer graphene, these theoretical findings regarding the systematic differences are fully consistent with the observations. 

Due to the uncertainties about the parameter values for the theoretical model and the approximations made in our approach, we cannot expect a fully quantitative description. However, the phenomenological input of requiring the single-layer to remain semi-metallic puts bounds on the bare interaction parameters that should be used in our model. The idea used in this paper is to scale down the ab-initio parameters for the interactions by an appropriate factor in order to compensate for the approximations, so as to render the fRG flow for the single layer free of divergences. With this remedy for the inexactness of our approach, the ABA$^*$ trilayer with remote hoppings remains semi-metallic and the energy scales of the unstable systems, namely the bilayer and the ABC trilayer come out in a quite realistic region below 30~meV. 

The type of the leading instability in these trilayer systems can also be read off from the fRG effective interactions near the instability. Here, for the interaction parameters determined by ab-initio methods\cite{wehling2011}, our calculations show a strong competition between SDW antiferromagnetic ordering and the interaction-driven `Kane-Mele' quantum spin Hall state. Which tendency wins depends on the detailed spatial profile of the interactions. A more longer-ranged behavior favors the QSH instability, and for pure onsite repulsions, the AF-SDW state is the clear winner. Both states would open up a bulk gap. The SDW states should have an interesting modulation of the ordered moments depending on the number of nearest neighbors, with smaller moments for higher coordination number\cite{lang2012}. The QSH state should be a true two-dimensional topological insulator for the trilayer case, as time-reversal invariant edge defects will not suffice to gap the three pairs of helical edge states completely, in contrast with the bilayer case (for a discussion of the edge-state robustness, see the review in  Ref.~\onlinecite{qizhang}). Hence, at least one pair of counter moving, spin-resolved helical edge states should survive time-reversal-invariant edge disorder and could hence serve as smoking gun for such a correlated ground state. This perspective is rather exciting, as this would be the first realization of the concept of an interaction-driven topological insulator ('topological Mott insulator')\cite{HonerkampRaghu2008}.

We acknowledge discussions with F. Zhang, T. O. Wehling, M.J. Schmidt, S. Bl\"ugel, E. Sasioglu, and financial support through the DFG research units FOR 723, 912, and 1162 and the DFG research training group GRK 1523. 


%
\end{document}